\documentclass[preprint,12pt]{elsarticle}  
\usepackage{amsmath}
\usepackage{amssymb}                                          
\usepackage{graphicx}

\usepackage{natbib}
\usepackage{mathtext}
\usepackage{epsf}
\usepackage{amsfonts}
\usepackage{gensymb}
\usepackage{float}

\usepackage[T2A]{fontenc}    
\usepackage[utf8]{inputenc}  

\begin{document}


\title{Detectability of neutron star --- white dwarf coalescences by eROSITA and ART-XC}

\author{A.D. Khokhriakova$^1$, S.B. Popov$^{1,2,3}$\\
$^1$ Department of Physics, Lomonosov Moscow State University\\
$^2$ Sternberg Astronomical Institute, Lomonosov Moscow State University\\
$^3$ Higher School of Economics, Moscow\\
}

\maketitle

\section*{Abstract}
    Coalescences of neutron stars and white dwarfs are relatively frequent phenomena, outnumbering other types of compact object mergers (neutron stars and black holes without involving white dwarfs) altogether. Such event potentially can produce not only optical, but also an X-ray burst. Transient source
    CDF-S XT2 \cite{2019Natur.568..198X} can be an example of this type of events as suggested by \cite{2019MNRAS.488..259F}. In this note we estimate the rate of these transients in the field of view of X-ray instruments on-board Spectrum-RG satellite. We demonstrate that during four years of the survey program several thousand 
    of events related to neutron star --- white dwarf mergers  might appear in the field of view of eROSITA. Collimation of X-ray emission can reduce this number. Smaller, but comparable number of transients is expected in the case of ART-XC telescope. However, due to relatively short duration --- $\lesssim 10^4$~s, --- mostly such transients might be visible just in one scan of telescopes ($\sim 40$~s), and so only a few photons are expected to be detected which makes definite identification without additional information very problematic.

\section{Introduction}

Coalescences of compact objects --- white dwarfs (WDs), neutron stars (NSs), and black holes (BHs), --- are powerful phenomena accompanied by different transients: gamma-ray bursts, gravitational wave emission, type Ia supernova, etc. 
Rate of different types of mergers varies from one event in few tens of years per a Milky Way-like galaxy for WD-WD 
(some fraction of such mergers --- $\sim15$\%, --- results in SN Ia which occur once in $\sim 200$~yrs per a Milky Way-like galaxy) \cite{2018MNRAS.476.2584M}
down to once in $\lesssim 10^5$~yrs per galaxy for NS-BH \cite{2019arXiv190704218P}.\footnote{Rates of compact object coalescence are often given in units [Gpc$^{-3}$~yr$^{-1}$]. Here we re-calculated them in units [per galaxy per year]. In \cite{2019arXiv190704218P} the authors give the value $\sim 10$~--~30~Gpc$^{-3}$~yr$^{-1}$. Rates of BH-BH and NS-NS coalescence are typically a little bit higher: up to $\sim 100$ Gpc$^{-3}$~yr$^{-1}$ \cite{2018arXiv180605820M, 2019arXiv190704218P} and up to $\sim300$ Gpc$^{-3}$~yr$^{-1}$ \cite{2017ApJ...846..170T}, respectively. }

Recently, NS-WD coalescences attracted much interest. As it was de\-mon\-strat\-ed by  \cite{2017MNRAS.467.3556B},  due to angular momentum losses via 
disk winds (in contrast with the models in which angular momentum is carried away by a jet), 
a wast majority of such events (with $M_\mathrm{WD}\gtrsim 0.2\, M_\odot$) proceeds in an unstable way on a very short time scale (from minutes to weeks depending on the WD mass, see Fig. 13 in \cite{2017MNRAS.467.3556B}). 
This rapid merger is necessarily accompanied by a huge energy release.
Thus, bright transients might appear at different wavelengths, including optical range.\footnote{Also gravitational wave signals are expected. They can be detected by eLISA and similar instruments sensitive to low-frequency waves.} 

In \cite{2019MNRAS.488..259F} the authors studied the outcome of NS-WD mergers in some details from the point of view of electromagnetic counterparts (see also a recent e-print \cite{2019arXiv190810866Z}). In particular, they suggested that due to jet formation a high-energy transient can appear. Moreover, it was proposed that the X-ray transient CDF-S XT2 discovered by \cite{2019Natur.568..198X} could be an example of such event. 

In \cite{2018A&A...619A..53T} the rate of NS-WD mergers was calculated for different ages of stellar populations. These results clearly demonstrate that this type of coa\-le\-sce\-nce outnumbers joint rate of NS-NS, NS-BH, and BH-BH mergers. 
Total (time intergrated) number of  NS-WD coalescence produced by a stellar population is: $(3-7)\times 10^{-5}$ per a solar mass of created stars. This corresponds to the rate $\sim (1-2)\times 10^{-4}$~yr$^{-1}$ in a Milky Way-like galaxy (it is important to underline, that mostly mergers happen within the first Gyr after binaries are formed). Note, that this value is $\lesssim10\%$ of the SN Ia rate.

In \cite{2017MNRAS.467.3556B} the authors also estimated the visibility of NS-WD merging systems in X-rays under assumption that the luminosity is equal to the Eddington value. Numbers are low, and it is not very probable that even a single event will be detected by eROSITA (extended ROentgen Survey with an Imaging Telescope Array) aboard Spectrum-RG satellite. However, as was shown in \cite{2017MNRAS.467.3556B}, this instrument can detect in relatively near-by galaxies many (from several tens to several hundreds) NS-WD systems with low-mass WDs evolving through the stage of stable mass transfer.
Here we explore a different possibility. 
We estimate the rate of eROSITA (and ART-XC) detections assuming that NS-WD coalescences can produce bright transients, similar to CDF XT2 \cite{2019Natur.568..198X}, due to jet (or another origin) X-ray emission.


\section{Estimate of the cosmic rate of NS-WD mergers}

In this section we estimate cosmic rate of NS-WD mergers integrated over different redshifts.

Hereafter, we assume flat Lambda-CDM cosmological model with the Planck 2018 cosmological parameters \cite{2018arXiv180706209P}: $H_0 = 67.4$ km s$^{-1}$ Mpc$^{-1}$, $\Omega_m = 0.315$, $\Omega_{\Lambda} = 0.685$. %

We use parameters of cosmic star formation history from \cite{2014ARA&A..52..415M}:
\begin{equation}
    \psi (z) = 0.015 \frac{(1+z)^{2.7}}{1 + [(1+z)/2.9]^{5.6}} \text{ M}_{\odot} \text{ yr}^{-1} \text{ Mpc}^{-3},
\end{equation}
where $z$ is redshift.
This function is the best fit for observational UV and IR data.

It is important to note, that our estimates below are sensitive only to the total star formation rate (SFR) at a given $z$, and do not depend on specific SFR in different types of galaxies, etc. 

The rate of NS-WD mergers is given by the equation:
\begin{equation}
    \frac{d\dot{N}_\mathrm{NSWD} (z)} {dz} = \frac{1}{1+z} \int_z^{z_{max}} \rho(z, z') \psi(z') \frac{dV_c(z')}{dz'} \frac{dt(z')}{dz'} dz'.
\end{equation}
Here $t$ is the age of the universe and we can write: 
\begin{equation}
    \frac{dt(z)}{dz} = - \frac{1}{H_0 (1+z) \sqrt{(1+z)^3 \Omega_m + \Omega_{\Lambda}}}.
\end{equation}
The maximum considered redshift $z_{max} = 11.247$ corresponds to the age of the universe 400 Myr, where intensive star formation began, 
$dV_\text{c}$ is differential comoving volume, and $\rho(z, z') = \rho(t(z) - t(z'))$ is determined by the  delay time function of NS-WD mergers. For the later quantity we use results obtained with the model $\alpha \alpha$ (with kick velocity distribution described in \cite{2017A&A...608A..57V}) from \cite{2018A&A...619A..53T}.

The $\alpha \alpha$-model is based on the classical energy balance of a common-envelope (CE) evolution:
\begin{equation}
    E_{\text{bin}} = \frac{G M_\text{d} M_\text{c}}{\lambda R} = \alpha_{\text{CE}} E_{\text{orbit}},
\end{equation}
where $M_\text{d}$ is the mass of the donor star, $M_\text{c}$ --- the mass of its core, $R$ --- its radius, $\lambda$ -- the structure parameter of its envelope, and $E_{\text{bin}}$ is the binding energy of the envelope. The orbital energy $E_{\text{orbit}}$ is considered as the source of energy to unbind the envelope with an efficiency determined by the parameter $\alpha_{\text{CE}}$.
The model assumes $\alpha_{\text{CE}} \lambda = 2$. Kick velocity distribution is taken from \cite{2017A&A...608A..57V}.

To calculate the differential comoving volume (and to take into account cosmological effects below) we use equations from \cite{1999astro.ph..5116H}:
\begin{equation}
    \frac{dV_c(z)}{dz} = \frac{c}{H_0} \frac{4 \pi D_M^2(z)}{\sqrt{(1+z)^3 \Omega_m + \Omega_{\Lambda}}}.
\end{equation}
Here $\Omega_m$ is the density of non-relativistic matter in units of the critical density, $\Omega_{\Lambda}$ is the density of the cosmological constant in units of the critical density, $D_\text{M}$ is comoving distance. 

The total rate of NS-WD mergers intergrated over all redshifts is $\sim$ 843 
per day for the whole sky. The maximum of the differential distribution of NS-WD merger rate is at $z = 1.8$.

Transient X-ray emission can be mildly collimated. We assume that jet emission fills solid angle  $\sim1$ steradian. Thus, the number of potentially observable transients is reduced by the factor $(1/4\pi)$ in comparison with the merger rate. The obtained distribution of transients rate is shown in Fig.~1.

\begin{figure}
\includegraphics[scale=0.9]{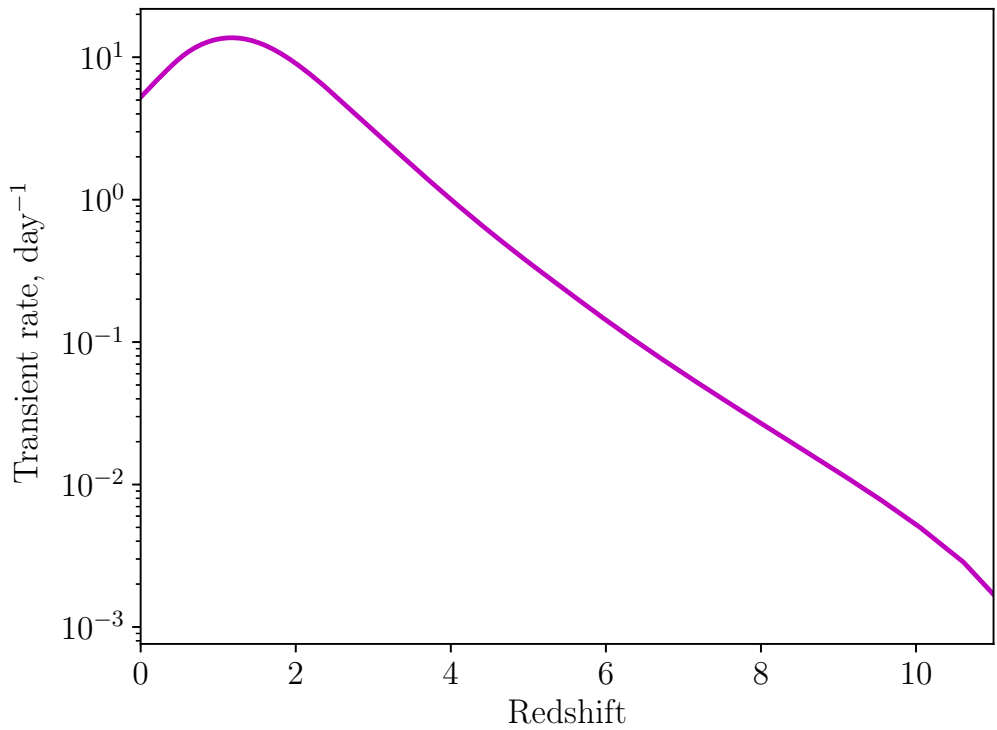}
\caption{The rate of NS-WD transients per day per unit redshift interval for different redshifts. It is assumed that collimated emission covers on a sphere the solid angle equal to one steradian. Total number of transients per day from all redshifts is equal to 67.} 
\end{figure}

\section{Log $N$ --- Log $S$ distribution for NS-WD mergers}

In \cite{2019MNRAS.488..259F} the authors obtain peak luminosities of transients $\lesssim 10^{47}$ erg s$^{-1}$. The peak isotropic luminosity of the CDF-S XT2 event was $L_\mathrm{X} \sim 3 \times 10^{45}$ erg s$^{-1}$. 
We construct the Log $N$ --- Log $S$ distribution for three characteristic luminosity values (Fig.~2) assuming that the luminosity distribution is a delta-function and that X-ray emission is collimated within a solid angle one steradian. The flux is given by the formula:
\begin{equation}
    F = \frac{L}{4 \pi D_\mathrm{L}^2},
\end{equation}
where $D_\mathrm{L} = (1 + z) D_\mathrm{M}$ is the luminosity distance. 
Spatial distribution of mergers is determined by the SFR from \cite{2014ARA&A..52..415M} and delay time between system's birth and NS-WD coalescence from \cite{2018A&A...619A..53T}.

In these estimates we neglect absorption and redshift influence (i.e., shift in spectral energy distribution) because the intrinsic spectrum of transients is unknown.

\begin{figure}
\includegraphics[scale=0.9]{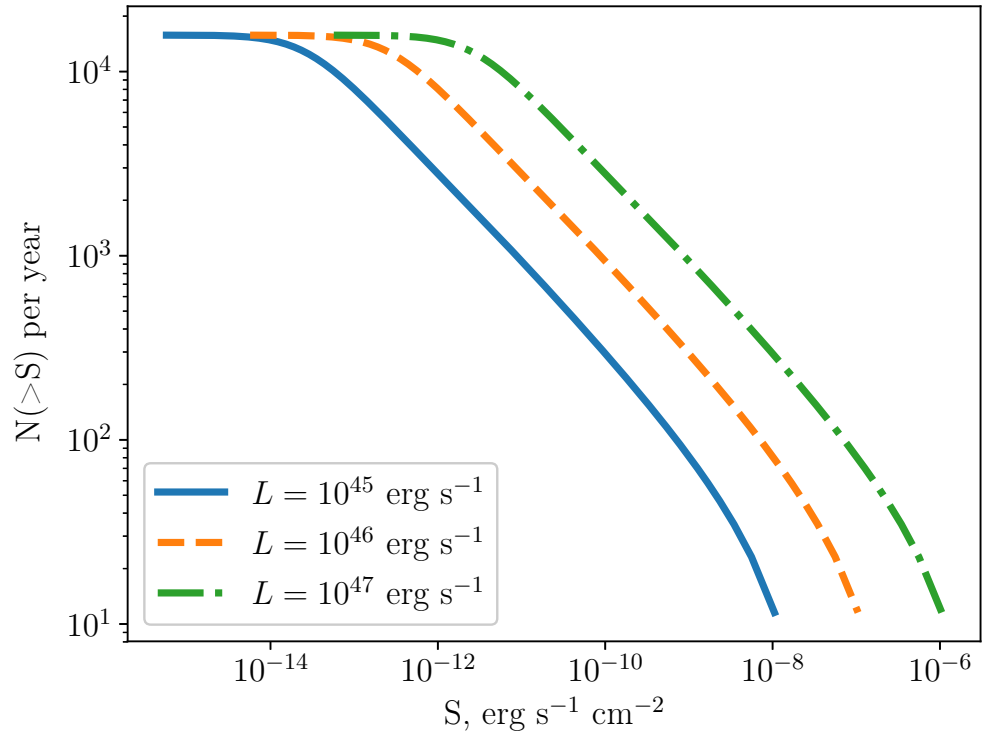}
\caption{All-sky Log $N$ --- Log $S$ distributions for different peak 
luminosities of NS-WD transients. S is the flux. The luminosity distribution is assumed to be a delta-function in each case. It is assumed that emission is collimated within one steradian. Flat parts of each curve correspond to $z> z_\mathrm{max}$.  } 
\end{figure}

\begin{figure}
\includegraphics[scale=0.9]{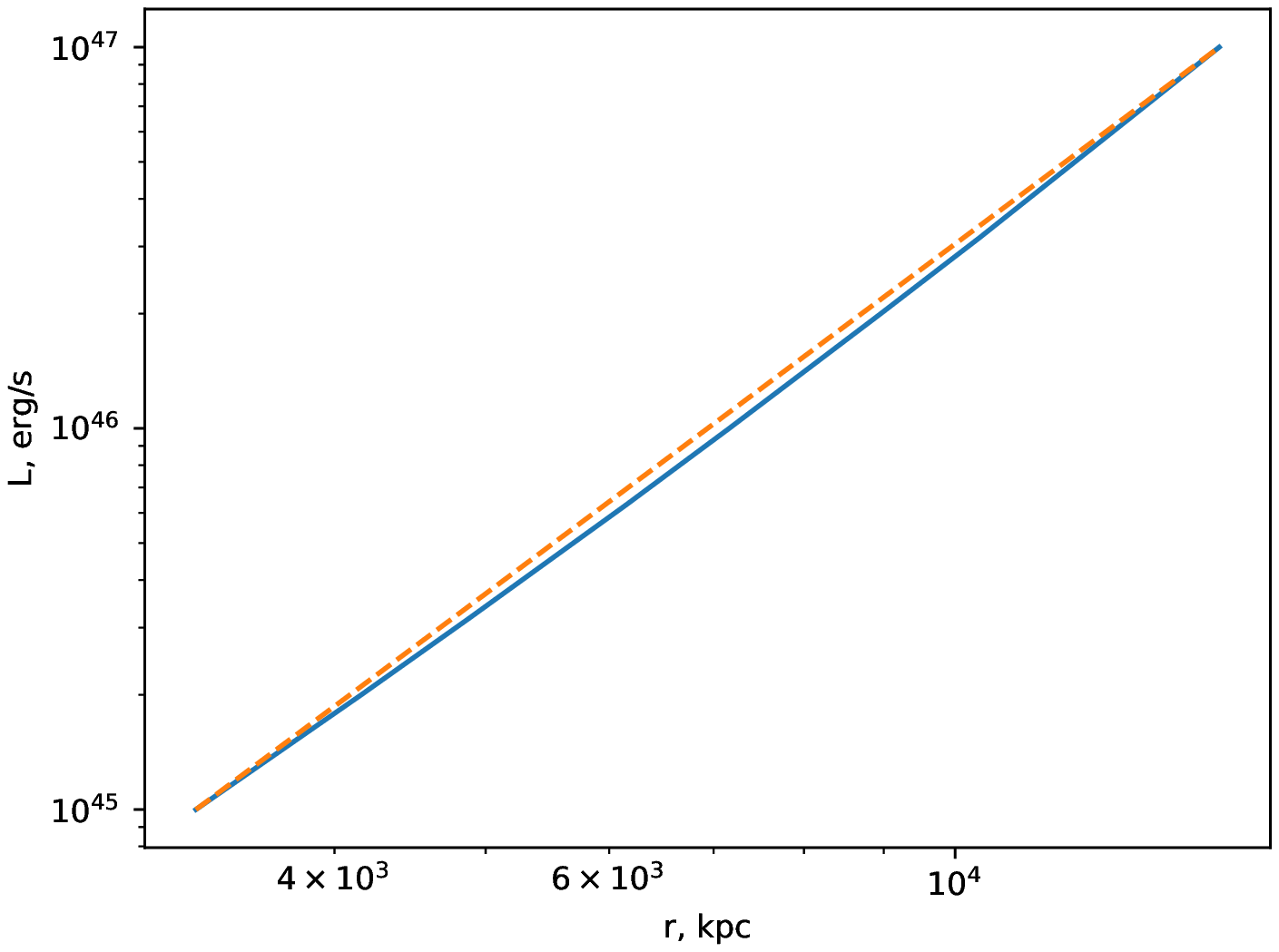}
\caption{Critical luminosity versus distance.  Above the critical luminosity an NS-WD merger with the spectrum similar to the one of CDF-S XT2 produces more than two counts at eROSITA. The solid line shows results of calculations.  The dashed line is a power-law approximation. }
\end{figure}

\section{Expected number of NS-WD transients in the field of view of eROSITA and ART-XC}

eROSITA  is the primary instrument of the Spectrum-Roentgen-Gamma (SRG) mission successfully launched in July 2019. 
The detailed description of the telescope is presented in \cite{merloni2012}.

The field of view (FoV) of eROSITA is 0.833 square degrees. 
During 4 years of survey observations it will perform 8 all-sky surveys in soft X-rays within the energy range  $\sim 0.3 - 10\, \mathrm{keV}$, completing one full circle on the sky (one scan) every 4 hours. 

In this section we estimate the number of NS-WD transients that fall in the field of view of eROSITA and ART-XC, the second telescope on-board SRG sensitive at the classical X-ray range from $\sim 3-5$~keV up to $\sim30$~keV.

The CDF-S XT2 X-ray outburst lasted for $\approx 20$ ks. However, after $t_\mathrm{break} = 2.3$ ks the flux started to decrease significantly. The $T_{90}$ parameter was estimated to be $\sim 11.1$ ks.
We therefore suggest that such transients last for $\sim 10$ ks.
During this time, telescope will cover a band of sky length $360^{\circ} \times { 10^4} / (4 \times 3600) = 250^{\circ}$.
This gives rate of 
\begin{equation}
N = N_{\text{NSWD}} \times \frac {\text{(eROSITA FoV)}} {4 \pi (\frac{180}{\pi})^2 \text{ deg}^2} \times 250 \text{ deg} = 4.3
\end{equation}
events per day falling in the eROSITA FoV.

Similarly, for the ART-XC telescope with FoV $ = 0.3$ square degrees, the rate will be $\sim 1.5$ events per day.

This values are obtained without beaming. If we account for this effect then numbers are correspondingly reduced. Thus, we expect that after 4 years of survey with $\sim 1$ steradian beaming we expect $\gtrsim 500$ 
events in the field of view of eROSITA, and $\gtrsim 180$ 
--- for ART-XC. 

\section{eROSITA rate of CDF-S XT2-like events}

In this section we model observations of CDF-S XT2-like events with eROSITA.

At the moment, there are no confirmed NS-WD coalescence detections. Accordingly, typical spectral parameters  of these events are unknown. However, in \cite{2019MNRAS.488..259F} the
authors claim that the X-ray transient CDF-S XT2 can be an example of such mergers. Therefore, it is reasonable to consider the spectrum of this event as a typical one in order to estimate how many objects eROSITA can observe and what are typical count rates.

In \cite{2019Natur.568..198X} the authors fit the spectrum of CDF-S XT2 with absorbed power-law with photon index $\Gamma = 1.57$ and hydrogen column density $N_\mathrm{H} = 7.7 \times 10^{21}$~cm$^{-2}$. Galactic contribution is fixed to a column density  $ 8.8 \times 10^{19}$~cm$^{-2}$.  


To calculate the number of photons that can be detected by eROSITA from the source with such spectrum, we use the method described in \cite{2019AstL...45..120K}.
In this approach we consider interstellar absorption (data taken from \cite{morrison1983}), dependence of the effective area of the telescope on wavelength (these data is taken from the web-site https://wiki.mpe.mpg.de/eRosita), and redshift influence (shift of the spectral energy distribution). Luminosity (and spectrum) are assumed to be constant during the whole burst duration, as according to observations \cite{2019Natur.568..198X} flux evolves very slowly --- as $\propto t^{-0.14}$, --- during the first several thousand of seconds.

We assume that the dependence of hydrogen column density on intergalactic distance is linear:
\begin{equation}
    N_H = 8.8 \times 10^{19}\text{ cm}^{-2} + 7.7 \times 10^{21}\text{ cm}^{-2} \frac{D_L}{4687 \text{ Mpc}}.
\end{equation}

The resulting number of detected photons is
\begin{equation}
N = \int_{E_1}^{E_2} {  \Delta t \text{ }\frac{f(E) e^{-\sigma N_\mathrm{H}} S_\mathrm{eff}(E) } {4 \pi D_\mathrm{L}^2}} dE,
\end{equation}
where $E$~is the photon energy in keV, $f(E)$ is the spectrum of the source, 
$ S_\mathrm{eff}(E)$ is the dependence of the effective area of the telescope on the wavelength, $\Delta t$ is the duration of the observation, $E_1$ and $E_2$ are the boundaries of the range of sensitivity of the telescope.   

The host galaxy of CDF-S XT2 is located at $z = 0.738$ (which corresponds to $D_\mathrm{L}=4687$~Mpc), and for the corresponding distance the number of photons detected by eROSITA is more than one hundred for the whole time of the burst. 
However, due to eROSITA observing strategy, such transient will be visible just in one scan of telescopes ($\sim 40$~s). In this case only 2 detected photons are expected. 
This value is above the expected background, which is $\sim 10^{-2}$ photons/s/pixel \cite{merloni2012}, and so the transient can be marginally detectable.

Similar calculations for ART-XC give $\lesssim  1.6 \times 10^{-2}$ photons/s for the actual response matrix (which corresponds to the effective area $\sim 400$~cm$^{-2}$ \cite{2019ExA....47....1P}). The background for this instrument is currently unknown, but it is expected to be larger than for eROSITA. Thus, we assume that ART-XC will not be able to detect CDF-S XT2-like events, unless they happen at significantly smaller distance.

Taking into account similar technical characteristics of {XMM-Newton} (in comparison with eROSITA) and long operation time in pointing mode, it is quite probable that archival data of this instrument fits better requirements for searching for X-ray transients related to NS-WD mergers.  

\section{Discussion}

\subsection{Uncertainties}

Degree of collimation of X-ray emission in transients under study is uncertain. Recently, in \cite{2019arXiv190710523D} the authors provided arguments that early X-ray afterglows after NS-NS coalescence can be unbeamed. Emission is provided by a newborn rapidly spinning magnetar. Such an object can be formed also in a NS-WD merger. Thus, similar arguments are applicable in this case, too. Then, the numbers given in Fig.~1 might be increased by the factor $4\pi$. 

In some of our estimates (Sec. 3) we neglected absorption in interstellar and intergalactic medium. As NS-WD mergers typically happen at least hundreds of million years after formation of stars, a binary can travel far from the birth site, so interstellar absorption in the host galaxy can be negligible. In principle, intergalactic contribution can be significant in the eROSITA range of sen\-si\-ti\-vi\-ty (for ART-XС it is less important). We can estimate it as 
$\frac{N_\mathrm{H}}{10^{22}\, \mathrm{cm}^{-2}}=\frac{DM}{333 \, \mathrm{pc\, cm}^{-3}}$, see \cite{2013ApJ...768...64H}, where $DM$ is dispersion measure, which for extragalactic sources (for example, Fast Radio Bursts --- FRBs) is about few hundreds for distances $\sim$ Gpc. Still, taking into account unknown spectral parameters (and other uncertainties) we prefer to make simple estimates with a toy model (see Fig. 2) in addition to calculations with parameters of CDF-S XT2 (see Sec. 5) which takes absorption into account. 

Uncertainties in SFR at large redshifts (see \cite{2014ARA&A..52..415M})
might not influence our results drastically, as according to \cite{2018A&A...619A..53T} 
the delay between binary formation and its coalescence is usually not very large: most of mergers happen after few Gyrs. Thus, the observable events might be dominated by sources born at $z\lesssim2$.
However, in these low redshifts the SFR is not known very precisely, too. And so, it is necessary to check how much this can influence our main results. 

We studied the effect of uncertainty in the SFR applying an alternative value  presented in a recent e-print \cite{2019arXiv190710523D}:
 \begin{equation}
     \mathrm{SFR}(z) \approx 0.25 \text{ } e^{-[\ln{((1+z)/3.16)]^2 / 0.524}} \text{ M}_\odot \text{ Mpc}^{-3} \text{ yr}^{-1}.
 \end{equation}
 With this rate we obtained a higher number of NS-WD mergers: 1983 
 per day across the sky assuming that emission is collimated within one steradian. This number might be compared with the rate of 843 events per day across the sky calculated above using the SFR given in eq.(1).
 This estimate makes the probability of detection of these transients with eROSITA (and/or ART-XC) much higher.

Uncertainties in the critical WD mass above which  transfer of matter to a NS is unstable are also not very important, as most of WDs have masses $>0.5$ M$_{\odot}$, signi\-fi\-cant\-ly above the recently derived limit $\gtrsim 0.2$ M$_\odot$, see \cite{2017MNRAS.467.3556B}.

The authors of \cite{2018A&A...619A..53T} made calculations only for the solar metallicity.
Of course, especially at large redshifts, metallicity can be different from this value. However, as we noticed above, the main contribution to the total rate is expected from sources born at $z \lesssim 2$,
where metallicity might be not very far from the solar value.  
We expect that uncertainties due to this factor might be smaller than those from unknown features of binary evolution (common envelope parameters, etc.) and NS properties (kick velocity distribution, in the first place).

\subsection{NS-WD mergers and FRBs}

Recently in \cite{2018Ap&SS.363..242L} NS-WD mergers were considered as progenitors of FRBs. Indeed, these collisions have several attractive features which make dis\-cus\-sion of this possibility interesting. Below we briefly comment on this proposal on the base of expected number of transients.

Total rate of NS-WD mergers obtained above is $\sim$ 850 per day across the sky (or  $\sim$ 2000 with another SFR).
This is noticeably lower than several thousands per day --- the estimated rate of FRBs if weak events are taken into account (see reviews in \cite{2018PhyU...61..965P, 2019A&ARv..27....4P, 2019arXiv190605878C}). 
Possible beaming reduces the number of observable transients related to NS-WD merger further (probably, by an order of magnitude). 

As all other catastrophic scenarios NS-WD mergers by themselves cannot explain repeating bursts \cite{2019arXiv190803507T}.
However, a magnetar formed after coalescence potentially can be discussed as a source of subsequent flares. In this case we expect that numerous repeating events differ from the first main burst, which can be accompanied by counterparts in other wavelengths \cite{2019MNRAS.488..259F}.
In particular, NS-WD coalescences can be detectable by future gravitational wave detectors and in optical surveys as transient events.

Thus, we conclude that coalescences of NSs and WDs cannot explain significant fraction of FRBs. However, it is not excluded that some small percentage of these events can be due to NS-WD interaction (for example, those which are not related to sites of recent star formation activity). At the moment it is not clear if the population of FRBs is uniform or not. Drastic increase of statistics of these events due to such instruments as CHIME, ASKAP, MeerKAT, Apertif, etc. expected in near future might help to resolve this issue. 

\section{Conclusions}

We estimated the total rate and distribution over redshift of NS-WD coalescences following calculations by \cite{2018A&A...619A..53T} and using starformation history from \cite{2014ARA&A..52..415M}. 
The total number of events is $\sim 800-2000$ per day depending on the assumed model of star formation history, and the maximum of the distribution is at $z\approx 1.8$. Basing on this calculations we claim that the rate of NS-WD mergers is too low to explain significant fraction of FRBs. 

We demonstrate that taking into account possible collimation of X-ray emission $\gtrsim 500$ events might appear in the field of view of eROSITA instrument aboard SRG satellite during 4 years of its survey program. Re\-spec\-ti\-ve\-ly, $\gtrsim 180$ are expected in the field of view of ART-XC --- the second instrument of the mission. Without collimation these numbers are an order of magnitude higher.

Expected numbers of counts are small because putative transients are expected to be visible just in one scan for tens of seconds, as the rotational period of the satellite is slightly longer than expected duration of a transient.

\section*{Acknowledgements}

We thank drs. Ekaterina Filippova and Galina Lipunova for useful discussions, and the referee for her/his comments.
The authors acknowledge the support from the Program of development of M.V. Lomonosov Moscow State University (Leading Scientific School ``Physics of stars, relativistic objects, and galaxies'').

\bibliography{bib}

\end{document}